# Efficient and accurate solution of wind-integrated optimal power flow based on enhanced second-order cone relaxation with rolling cutting plane technique

Zhaojun Ruan [a], Botao Gao [a], Libao Shi [a]

[a] Institute for Ocean Engineering, Tsinghua Shenzhen International Graduate School, Tsinghua University, Shenzhen, P. R. China

## Abstract

The integration of large-scale renewable energy sources, such as wind power, poses significant challenges for the optimal operation of power systems owing to their inherent uncertainties. This paper proposes a solution framework for wind-integrated optimal power flow (OPF) that leverages an enhanced second-order cone relaxation (SOCR), supported by a rolling cutting plane technique. Initially, the wind generation cost, arising from discrepancies between scheduled and actual wind power outputs, is meticulously modeled using a Gaussian mixture model based on historical wind power data. This modelled wind generation cost is subsequently incorporated into the objective function of the conventional OPF problem. To achieve the efficient and accurate solution for the wind-integrated OPF, effectively managing the constraints associated with AC power flow equations is essential. In this regard, a SOCR, combined with a second-order Taylor series expansion, is employed to facilitate the convex approximation of the AC power flow equations. Additionally, a warm-start strategy, grounded in a proposed rolling cutting plane technique, is devised to reduce relaxation errors and enhance computational efficiency. Finally, the effectiveness and efficiency of the proposed solution framework are demonstrated across various case studies. Specifically, the influence of wind power cost is also examined, further highlighting the practical implications of the proposed solution framework.

## 1. Introduction

Power flow analysis is fundamental to power system research, as it forms the basis for effective planning, operational management, and real-time control strategies within power systems. Building upon these foundational studies, Carpentier's groundbreaking introduction of the optimal power flow (OPF) framework in 1962 marked a significant advancement in achieving operational optimization and economic efficiency [1]. Specifically, the OPF problem is formulated to minimize the total costs associated with dispatching active and reactive power from multiple generating units while simultaneously fulfilling demand requirements at various buses. To accomplish these objectives, the OPF framework must adhere to essential physical and technical constraints, including those dictated by Ohm's and Kirchhoff's Laws, along with the thermal limits of transmission lines (branch flow limits) [2]. By integrating economic and engineering considerations, the OPF paradigm not only ensures cost-effective system operation but also provides a robust mechanism to enhance network reliability and resource utilization in modern power grids. With the increasing penetration of renewable energy sources, particularly wind power, significant challenges have emerged for conventional power grid operations due to the stochastic and volatile nature of wind generation. Traditional OPF formulations typically focus on minimizing fossil fuel costs while generally neglecting the cost implications associated with wind power integration. This oversight highlights the critical need for the development of modern OPF models that explicitly quantify wind power costs within contemporary power systems. Addressing the wind-integrated OPF problem requires the simultaneous resolution of two fundamental challenges: First, the inherent non-convex nature of the OPF problem necessitates advanced mathematical treatment to ensure solution feasibility. Second, there exists a pressing need to devise robust methodologies for the quantification of wind generation costs, particularly given its intermittent characteristics and the associated uncertainties.

However, the most significant nonlinear and non-convex factors within the OPF problem arise from AC power flow equations, which inherently reflect the nonlinear coupling between power and voltage [3]. Theoretically, finding a global optimal solution to the AC OPF problem is classified as NP-hard [4]. In large-scale systems, the strong non-convex and nonlinear characteristics of AC OPF result in considerable computational costs and challenges in guaranteeing convergence. Consequently, reducing computational complexity while enhancing solution quality remain prime motivations for researchers. In recent years, there has been a heightened emphasis on AC OPF problems, driven by advancements in power system analysis and optimization theories [4]. The diverse approaches to solving AC OPF can be broadly categorized into three classes: (i) exact methods [5] - [7], (ii) convex relaxations [8] - [12], and (iii) linear approximations [13] - [24].

Exact methods aim to preserve the physical significance of the non-convex AC power flow equations, thus producing either local or global optimal solutions. Techniques such as primal-dual nonlinear interior point methods (IPMs) [5], sequential quadratic programming [6], and sequential linear programming [7] represent notable solutions. While capable of yielding high-quality local solutions, these methods often incur substantial computational costs for large-scale systems and cannot guarantee solution quality or global optimality [2].

To enhance computational efficiency and ensure convergence, convex relaxation techniques for AC OPF have attracted considerable interest. These methods relax the non-convexity of AC OPF via applying convex formulations to power flow equality constraints, exemplified by second-order cone relaxation (SOCR) [8, 9], quadratic convex relaxation [10], semi-definite relaxation [11], and the construction of convex envelopes or hulls. Convex relaxations allow for polynomial time solutions to OPF and provide a lower bound for the original problem's solution [12]. Nonetheless, these methods are not always exact, when a non-zero duality gap occurs, recovering AC-feasible solutions can be challenging, and pre-determining the optimal gap for different systems is not feasible [4]. This lack of guaranteed accuracy limits their practical application in the power industry.

Since its inception, DC power flow has emerged as one of the most prevalent linear approximations in power system analyses owing to its simplicity [13]. This approach assumes constant voltage magnitudes and negligible reactive power impacts, transforming the network into a set of linear equations concerning bus voltage angles and active power injections [14]. Although the DC OPF model efficiently supports sensitivity analysis, electricity market clearing, and power system dispatch, it oversimplifies voltage magnitude and reactive power, potentially yielding solutions that may be infeasible under more realistic operating conditions [15]. Linear approximation models, based on mathematical approximations, cannot guarantee the optimality of the original AC OPF but can still offer high-quality solutions with robust performance and reduced computational requirements [16]. These approaches generally approximate the nonlinear power flow equations around a specified operating point, often employing Taylor series expansions [17]. The accuracy of these linearized frameworks depends on how much the system's actual operating point deviates from the initial approximation [18]. Many existing linear approximation methods utilize Taylor series expansions combined with mathematical substitutions. For instance, *Zhang et al.* [19] employed both voltage magnitude and voltage angle as independent variables, achieving an acceptable linearization error by leveraging the quasi-linear relationship between bus voltage angles and branch active power. Additionally, piecewise linearization techniques were introduced to handle network loss approximations. However, significant departures of voltage magnitude from nominal values or a small X/R ratio can lead to substantial approximation errors [20]. To address this issue, *Xiao et al.* [21] employed second-order Maclaurin series to enhance accuracy in branch power flow and network loss calculations, while *Yang et al.* [16] applied first-order Taylor series expansions alongside warm-start methods to refine solution quality. A logarithmic transformation of voltage magnitude was also proposed to better capture the nonlinear impact of voltage on both active and reactive power flows as well as network losses [22]. Additionally, investigations leveraging the orthogonality properties of Legendre polynomials were conducted to minimize the discrepancies between linear approximations and exact AC OPF solutions [23]. Under relatively stable operating conditions, the linear approximation error is generally acceptable. However, over a broader operating range, the state variables in linearized power flow models often deviate substantially from the initial approximation point, leading to a market increase in linearization error [24]. Traditional approaches based on a flat start and Maclaurin series expansions fail to accurately capture the true operating point. In contrast, employing warm-start iterative method can effectively reduce the linear approximation error and yield more precise insights regarding the system's operating conditions [16].

Reasonable quantification of wind power costs is essential for developing viable OPF frameworks in wind-integrated power systems. Current research predominantly confines the stochastic and volatile characteristics of wind generation to constraint formulations, establishing optimization paradigms based on robust [25, 26] or stochastic programming [27, 28] methodologies. While some studies [29, 30] have attempted to incorporate wind-related cost components into objective functions, their economic models frequently suffer from inadequate representation of practical operational factors, particularly with respect to uncertainty quantification metrics and real-world cost allocation mechanisms. Building upon the foundational work discussed in [31], this paper leverages a data-driven approach for wind generation cost modeling using Gaussian mixture model (GMM). By leveraging the integral invariance property inherent to GMM, the corresponding analytical expressions are derived to quantify the associated costs under both wind power shortage and surplus scenarios. To further enhance both accuracy and efficiency of solving the wind-integrated OPF problem, this paper proposes an enhanced SOCR framework. First, a second-order Taylor series expansion is applied to formulate an iterative version of the power flow equations. The bilinear terms introduced by the second-order Taylor series expansion are subsequently treated via SOCR, and a warm-start strategy with rolling cutting-plane technique is proposed to progressively tighten the associated relaxation gap. Finally, the effectiveness and efficiency of the proposed solution framework are validated through simulations on test systems of varying sizes.

The remainder of this paper is organized as follows: Section II provides a comprehensive exposition of the proposed wind-integrated OPF model with enhanced SOCR framework, including the modeling of wind generation costs. Section III elaborates on the rolling cutting-plane technique and the warm-start strategy for the proposed wind-integrated OPF model. Section IV presents detailed numerical simulations and comparative analysis of experimental results. Finally, the conclusions are summarized in Section V.

## 2 Problem Formulation

In this study, an efficient and accurate solution of wind-integrated OPF based on enhanced SOCR with rolling cutting plane technique is elaborately discussed. First, a GMM is employed to characterize wind power output distributions based on historical data, subsequently quantifying costs through the integral invariance property of the GMM. This methodology ultimately yields a closed-form analytical expression for wind generation cost, which is integrated into the objective function of OPF problem. Next, building upon the original AC OPF formulation, the power flow equations in polar coordinates are reformulated using second-order Taylor series expansion, and the SOCR is introduced to address the non-convex terms. Additionally, the specific conditions under which the convex relaxation is exact are discussed, and a linear relaxation of the branch flow limits is performed to further reduce the computational complexity of the model. Finally, a warm-start strategy leveraging the proposed rolling cutting plane technique is devised to solve the wind-integrated OPF problem accurately and efficiently. The overall research framework is depicted in Fig. 1.

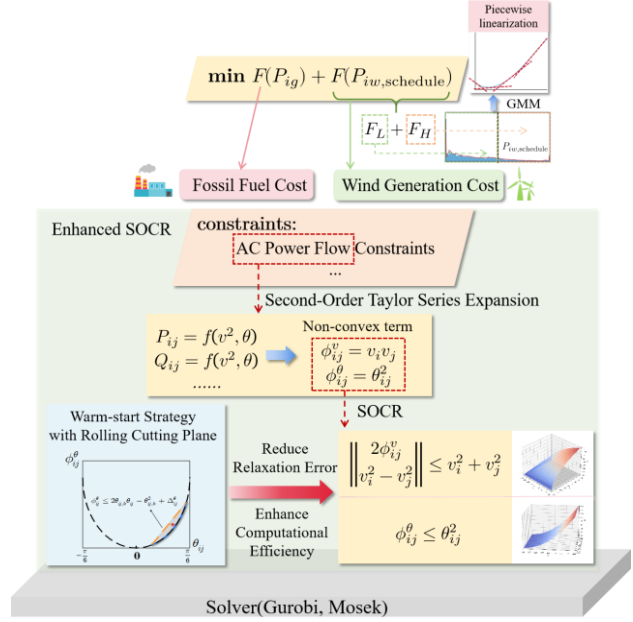

Fig. 1 Schematic diagram of overall research framework

### 2.1 Data-driven approach for modeling wind generation cost

To enhance the rationality of wind generation cost modeling, a non-parametric estimation framework, specifically the GMM, is utilized to characterize historical distribution patterns of wind power output. Following this, the methodology systematically evaluates both under-generation costs (associated with insufficient system reserve capacity) and over-generation costs (linked to wind curtailment policies) using an integrated risk assessment framework. These cost components are analytically expressed as follows:

$$F_\text{L} = k_\text{L} \cdot \Pr(P_\text{w} < P_\text{schedule}) \cdot (P_\text{schedule} - \text{E}_\text{L}(P_\text{w})) \tag{1a}$$

$$F_\text{H} = k_\text{H} \cdot \Pr(P_\text{w} > P_\text{schedule}) \cdot (\text{E}_\text{H}(P_\text{w}) - P_\text{schedule}) \tag{1b}$$

$$\Pr(P_\text{w} < P_\text{schedule}) = \int_{-\infty}^{P_\text{schedule}} f(P_\text{w}) \, dP_\text{w} = \Phi(P_\text{w}) = \sum_{k=1}^{K} \omega_k \left[ \int_{P_\text{w} \leq P_\text{schedule}} \mathcal{N}_k(P_\text{w}|\mu_k, \sigma_k) dP_\text{w} \right] \tag{1c}$$

$$\Pr(P_\text{w} > P_\text{schedule}) = 1 - \Phi(P_\text{w}) \tag{1d}$$

$$\text{E}_\text{L}(P_\text{w}) = \frac{\int_0^{P_\text{schedule}} f(P_\text{w}) \cdot P_\text{w} dP_\text{w}}{\Pr(P_\text{w} < P_\text{schedule})} = \frac{\sum_{k=1}^{K} \omega_k \{\sigma_k^2[\mathcal{N}_k(0) - \mathcal{N}_k(P_\text{schedule})] + \mu_k[\Phi_k(P_\text{schedule}) - \Phi_k(0)]\}}{\sum_{k=1}^{K} \omega_k \left[ \int_{P_\text{w} \leq P_\text{schedule}} \mathcal{N}_k(P_\text{w}|\mu_k, \sigma_k) dP_\text{w} \right]} \tag{1e}$$

$$\text{E}_\text{H}(P_\text{w}) = \frac{\int_{P_\text{schedule}}^{P_\text{max}} f(P_\text{w}) \cdot P_\text{w} dP_\text{w}}{\Pr(P_\text{w} > P_\text{schedule})} = \frac{\sum_{k=1}^{K} \omega_k \{\sigma_k^2[\mathcal{N}_k(P_\text{schedule}) - \mathcal{N}_k(P_\text{max})] + \mu_k[\Phi_k(P_\text{max}) - \Phi_k(P_\text{schedule})]\}}{1 - \sum_{k=1}^{K} \omega_k \left[ \int_{P_\text{w} \leq P_\text{schedule}} \mathcal{N}_k(P_\text{w}|\mu_k, \sigma_k) dP_\text{w} \right]} \tag{1f}$$

where $\mathcal{N}(x|\mu_k, \sigma_k)$ denotes the Gaussian probability density function (PDF) with mean $\mu_k$ and variance $\sigma_k$, and $\Phi(\cdot)$ represents the corresponding cumulative distribution function (CDF). The cost coefficients $k_L$ and $k_H$ (\$/MWh) quantify the economic penalties for wind power shortage and surplus, respectively. $\Pr(P_w < P_{schedule})$ and $\Pr(P_w > P_{schedule})$ are defined as the probabilities of wind power shortage and surplus, respectively. The conditional expected power levels (in MW) are given by $E_L(P_w)$ and $E_H(P_w)$. For the detailed derivation process, please refer to Appendix.

In accordance with national energy policy regulations, onshore wind generation adheres to the local benchmark coal-fired power price [32], with the base rate ranging between \$48~62/MWh. The penalty coefficients are set as $k_H = \$50/\text{MWh}$, $k_L = \$60/\text{MWh}$. As illustrated in Fig. 2, the cost characteristics of wind generation exhibit significant variation with scheduled output adjustments.

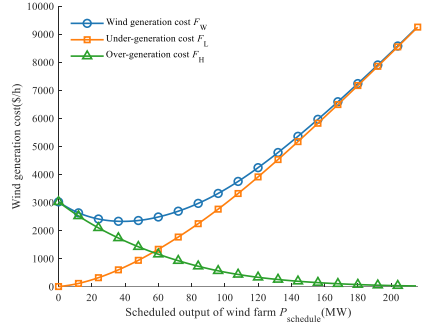

Fig. 2. Wind generation cost versus scheduled output of wind farm

From Fig. 2, as the scheduled output of wind farm increases, the system experiences a progressively higher probability of wind power shortage accompanied by a diminishing probability of power surplus. Consequently, the operational cost associated with power shortage demonstrates a monotonic upward trend, while the penalty cost related to power surplus shows a continuous downward tendency. This phenomenon leads to the costs of wind generation integration exhibiting a characteristic convex pattern, with an initial decline followed by subsequent growth.

As shown in Fig. 3, the costs of wind generation integration demonstrate significant sensitivity to variations in cost coefficients.

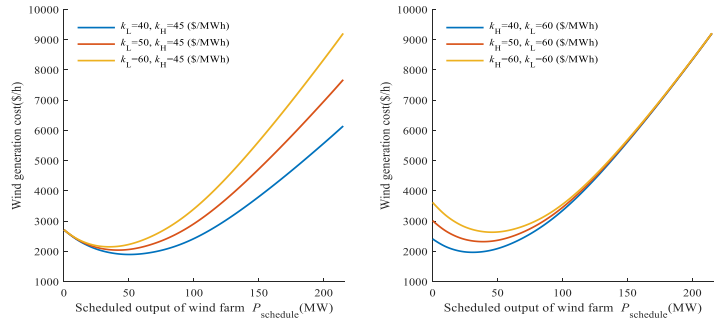

Fig. 3. Wind generation cost versus scheduled wind farm power output with $k_H$ and $k_L$

When analyzing the impact of cost coefficient variations (with $k_H$ maintained constant at \$45/MWh in the left panel), the escalation of $k_L$ magnifies the operational impact of wind power shortage. This condition reflects insufficient system reserve capacity, compelling dispatch strategies to prioritize system security, thereby resulting in a 8.8% reduction in optimal scheduled wind power output per 25% increase in $k_L$.

Conversely, the right panel analysis (with $k_L$ fixed at \$60/MWh) reveals that heightened $k_H$ values amplify the influence of wind power surplus on dispatch decisions. Under this scenario, the system exhibits an increased propensity for renewable energy accommodation, with dispatch optimization emphasizing environmental considerations. Consequently, the optimal scheduled wind power output escalates by approximately 8.2% for every 25% increment in $k_\mathrm{H}$.

## 2.2 Original AC OPF Formulation

Consider a standard $n$-bus power network $\mathcal{G} = (\mathcal{N}, \mathcal{E})$ with $ng$ generators and $nw$ wind turbine generators, where $\mathcal{N}$ denotes the set of all buses. Here, $\mathcal{E}$ and $\mathcal{E}_i$ represent the set of all branches and the set of branches connected to bus $i$, respectively. In this paper, two typical power system topologies are examined: *radial* and *meshed*. Medium and low voltage distribution networks are generally radial, whereas high voltage transmission networks are typically meshed. The complex voltage $\tilde{v}_i$ at bus $i$ is denoted by $\tilde{v}_i = v_i e^{\mathbf{i}\theta_i} = v_i \cos\theta_i + \mathbf{i}v_i \sin\theta_i = v_{i,d} + \mathbf{i}v_{i,q}$, where $\mathbf{i} = \sqrt{-1}$. The power flow equations are formulated in polar coordinates in this paper. Transmission lines and phase-shifting transformers are modeled using the $\pi$-model, with series admittance $\tilde{y}_{ij} = g_{ij} + \mathbf{i}b_{ij}$, where $g_{ij}$ and $b_{ij}$ are the conductance and susceptance in branch $ij$, respectively. Additionally, let $b_{ij}^{\mathrm{ch}}$ denote the total charging susceptance of branch $ij$, and $g_i^{\mathrm{sh}}$ and $b_i^{\mathrm{sh}}$ represent the conductance and susceptance of the shunt element at bus $i$. The complex tap ratio of a phase-shifting transformer is given by $\tilde{t}_{ij} = \tau_{ij} e^{\mathbf{i}\varphi_{ij}^{\mathrm{shift}}}$, where for a transmission line, $\tau_{ij} = 1$ and $\varphi_{ij}^{\mathrm{shift}} = 0$. Further, the notations $\mathfrak{R}(\cdot)$ and $\mathfrak{I}(\cdot)$ represent the real and imaginary parts, respectively. The notations $\overline{(\cdot)}$ and $\underline{(\cdot)}$ denote the upper and lower bounds of a variable, and $(\cdot)^*$ denotes the complex conjugation.

In the original AC OPF problem, the formulation is typically expressed as an optimization problem. Recent studies have adopted the bus injection model (BIM) and the branch flow model to represent the power flow equations. Within the BIM framework, the voltage-based formulations are utilized. To explicitly separate the real and imaginary parts, voltage-based formulations can be categorized into rectangular and polar coordinate representations. For the subsequent derivations, the polar coordinate form is employed. Consequently, the AC OPF problem can be expressed as:

$$\underset{P_{ig}, Q_{ig}, v_i, \theta_i}{\text{minimize}} \sum_{ig \in \mathcal{N}_{ng}} f_{ig}(P_{ig}) \tag{2a}$$

subject to

$$\underline{P}_{ig} \leq P_{ig} \leq \overline{P}_{ig}, \ \underline{Q}_{ig} \leq Q_{ig} \leq \overline{Q}_{ig}, \qquad \forall ig \in \mathcal{N}_{ng} \tag{2b}$$

$$\underline{v}_i \leq v_i \leq \overline{v}_i, \ \underline{\theta}_{ij} \leq \theta_i - \theta_j \leq \overline{\theta}_{ij}, \qquad \forall i \in \mathcal{N}, \forall ij \in \mathcal{E} \tag{2c}$$

$$\sum_{ig \in \mathcal{N}_{ng}} P_{ig} - P_{iD} - g_i^{\mathrm{sh}} v_i^2 = \sum_{ij \in \mathcal{E}_i} P_{ij} + \sum_{ji \in \mathcal{E}_i} P_{ji}, \qquad \forall i \in \mathcal{N} \tag{2d}$$

$$\sum_{ig \in \mathcal{N}_{ng}} Q_{ig} - Q_{iD} + b_i^{\mathrm{sh}} v_i^2 = \sum_{ij \in \mathcal{E}_i} Q_{ij} + \sum_{ji \in \mathcal{E}_i} Q_{ji}, \qquad \forall i \in \mathcal{N} \tag{2e}$$

$$P_{ij} = g_{ij}^f v_i^2 - g_{ij}^c v_i v_j \cos\theta_{ij} - b_{ij}^c v_i v_j \sin\theta_{ij}, \qquad \forall i,j \in \mathcal{N}, \forall ij \in \mathcal{E} \tag{2f}$$

$$P_{ji} = g_{ij}^t v_j^2 - g_{ji}^c v_j v_i \cos\theta_{ji} - b_{ji}^c v_j v_i \sin\theta_{ji}, \qquad \forall i,j \in \mathcal{N}, \forall ji \in \mathcal{E} \tag{2g}$$

$$Q_{ij} = -b_{ij}^f v_i^2 + b_{ij}^c v_i v_j \cos\theta_{ij} - g_{ij}^c v_i v_j \sin\theta_{ij}, \qquad \forall i,j \in \mathcal{N}, \forall ij \in \mathcal{E} \tag{2h}$$

$$Q_{ji} = -b_{ij}^t v_j^2 + b_{ji}^c v_j v_i \cos\theta_{ji} - g_{ji}^c v_j v_i \sin\theta_{ji}, \qquad \forall i,j \in \mathcal{N}, \forall ji \in \mathcal{E} \tag{2i}$$

$$P_{ij}^2 + Q_{ij}^2 \leq \overline{S}_{ij}^2, P_{ji}^2 + Q_{ji}^2 \leq \overline{S}_{ji}^2, \qquad \forall ij, ji \in \mathcal{E} \tag{2j}$$

where $g_{ij}^f = \mathfrak{R}[(\tilde{y}_{ij}^* - 0.5\mathbf{i}b_{ij}^{\mathrm{ch}})/|\tilde{t}_{ij}|^2]$, $g_{ij}^t = \mathfrak{R}(\tilde{y}_{ij}^* - 0.5\mathbf{i}b_{ij}^{\mathrm{ch}})$, $b_{ij}^f = -\mathfrak{I}[(\tilde{y}_{ij}^* - 0.5\mathbf{i}b_{ij}^{\mathrm{ch}})/|\tilde{t}_{ij}|^2]$, $b_{ij}^t = -\mathfrak{I}(\tilde{y}_{ij}^* - 0.5\mathbf{i}b_{ij}^{\mathrm{ch}})$, $g_{ij}^c = \mathfrak{R}(\tilde{y}_{ij}^*/\tilde{t}_{ij})$, $b_{ij}^c = -\mathfrak{I}(\tilde{y}_{ij}^*/\tilde{t}_{ij})$, $g_{ji}^c = \mathfrak{R}(\tilde{y}_{ij}^*/\tilde{t}_{ij}^*)$, and $b_{ji}^c = -\mathfrak{I}(\tilde{y}_{ij}^*/\tilde{t}_{ij}^*)$; $P_{ig}$ and $Q_{ig}$ denote the active and reactive power outputs of the unit $i$; $P_{iD}$ and $Q_{iD}$ denote the load demand at bus $i$; $v_i$ and $\theta_i$ represent the voltage magnitude and voltage angle; $P_{ij}$, $P_{ji}$, $Q_{ij}$, and $Q_{ji}$ denote the active and reactive power flows on the branches $ij(ji)$; $\overline{S}_{ij(ji)}$ denotes the upper bound of the apparent power magnitude. The objective function, as defined in (2a), is typically formulated to minimize the fossil fuel cost, expressed in the quadratic form $f_{ig}(P_{ig}) = c_{2,ig}(P_{ig})^2 + c_{1,ig}(P_{ig}) + c_{0,ig}$, in which

$c_{\kappa,ig}(\kappa \in \{0,1,2\})$ represents the cost coefficients. For nonlinear or other non-convex objective functions, a piecewise linearization approach can be adopted to achieve linearity. In the formulation, constraints (2b) represent the active and reactive power output limits of the generators, respectively. Constraints (2c) enforce the bus voltage magnitude limits and the branch angle difference limits. Constraints (2d) and (2e) ensure the active and reactive power balance at the buses. Constraints (2f) - (2i) govern the active and reactive power flows on the branches. Finally, constraints (2j) specify the branch flow limits.

Owing to the presence of strongly non-convex terms in constraints (2f)–(2i), such as trigonometric functions and trilinear expressions, the AC OPF problem is inherently challenging to solve. A primary purpose of this study is to reduce the model's non-convexity and nonlinearity as much as possible while preserving the required solution accuracy.

## 2.3 Second-order Taylor series expansion of power flow equations

Compared with conventional DC approximation methods, the method presented in this paper retains quadratic terms in the power flow equations, thereby more accurately capturing the nonlinear characteristics of network losses and enhancing the approximation accuracy of the equations. Additionally, traditional cold-start linearized power flow models typically assume that the voltage magnitude at each bus in the transmission network is approximately 1.0 p.u. (i.e., $v_i = 1$) and that the phase angle differences across transmission lines are negligible (i.e., $\theta_{ij} = 0$). However, these assumptions are often unrealistic in practical power system operations. To address this, a warm-start approach that iteratively corrects any resultant errors is proposed in this study. Assuming a predetermined initial operating point $(v_{i,k}, \theta_{ij,k})$, the second-order Taylor series expansion of trigonometric functions in polar coordinates can be formulated as follows:

$$\sin\theta_{ij} \approx \alpha_{ij,k}^1 \theta_{ij} + \alpha_{ij,k}^0, \ \cos\theta_{ij} \approx \beta_{ij,k}^2 \theta_{ij}^2 + \beta_{ij,k}^1 \theta_{ij} + \beta_{ij,k}^0 \tag{3a}$$

where

$$\alpha_{ij}^1 = \cos\theta_{ij,k}, \ \alpha_{ij}^0 = \sin\theta_{ij,k} - \theta_{ij,k}\cos\theta_{ij,k} \tag{3b}$$

$$\beta_{ij}^2 = -\frac{1}{2}\cos\theta_{ij,k}, \ \beta_{ij}^1 = -\sin\theta_{ij,k} + \theta_{ij,k}\cos\theta_{ij,k}, \beta_{ij}^0 = \cos\theta_{ij,k} + \theta_{ij,k}\sin\theta_{ij,k} - \frac{1}{2}\theta_{ij,k}^2\cos\theta_{ij,k} \tag{3c}$$

In (3b) and (3c), all parameters are constants. Given that (2g) and (2i) have a similar form to (2f) and (2h), and without loss of generality, substituting (3a) into (2f) and (2h) yields the following results:

$$P_{ij,k} = g_{ij}^f v_i^2 - g_{ij}^c v_i v_j \left(\beta_{ij,k}^2 \theta_{ij}^2 + \beta_{ij,k}^1 \theta_{ij} + \beta_{ij,k}^0\right) - b_{ij}^c v_i v_j \left(\alpha_{ij,k}^1 \theta_{ij} + \alpha_{ij,k}^0\right) \tag{3d}$$

$$Q_{ij,k} = -b_{ij}^f v_i^2 + b_{ij}^c v_i v_j \left(\beta_{ij,k}^2 \theta_{ij}^2 + \beta_{ij,k}^1 \theta_{ij} + \beta_{ij,k}^0\right) - g_{ij}^c v_i v_j \left(\alpha_{ij,k}^1 \theta_{ij} + \alpha_{ij,k}^0\right) \tag{3e}$$

Since the variables $v$ and $\theta$ in (3d) and (3e) remain tightly coupled, strong non-convexity still exists. Therefore, based on the specified operating point, a first-order Taylor series expansion is performed as follows:

$$v_i v_j \theta_{ij} \approx v_{i,k} v_{j,k} \theta_{ij} + \left(v_i v_j - v_{i,k} v_{j,k}\right)\theta_{ij,k}, \ v_i v_j \theta_{ij}^2 \approx v_{i,k} v_{j,k} \theta_{ij}^2 + \left(v_i v_j - v_{i,k} v_{j,k}\right)\theta_{ij,k}^2 \tag{3f}$$

By substituting (3f) into (3d) and (3e), respectively, and eliminating the coupling terms, the power flow equations that incorporating both bilinear terms and quadratic angle terms can be expressed as follows:

$$P_{ij,k} = g_{ij}^f v_i^2 - g_{ij,k}^P v_i v_j - b_{ij,k}^P \left(\theta_{ij} - \theta_{ij,k}\right) - b_{ij,k}^{Ploss} \left(\theta_{ij}^2 - \theta_{ij,k}^2\right) \tag{3i}$$

$$Q_{ij,k} = -b_{ij}^f v_i^2 + b_{ij,k}^Q v_i v_j - g_{ij,k}^Q \left(\theta_{ij} - \theta_{ij,k}\right) - g_{ij,k}^{Qloss} \left(\theta_{ij}^2 - \theta_{ij,k}^2\right) \tag{3j}$$

where

$$g_{ij}^P = \left(g_{ij}^c \beta_{ij,k}^0 + b_{ij}^c \alpha_{ij,k}^0\right) + \left(g_{ij}^c \beta_{ij,k}^1 + b_{ij}^c \alpha_{ij,k}^1\right)\theta_{ij,k} + g_{ij}^c \beta_{ij,k}^2 \theta_{ij,k}^2 = g_{ij}^c \cos\theta_{ij,k} + b_{ij}^c \sin\theta_{ij,k} \tag{3k}$$

$$b_{ij}^P = \left(g_{ij}^c \beta_{ij,k}^1 + b_{ij}^c \alpha_{ij,k}^1\right) v_{i,k} v_{j,k} = \left(-g_{ij}^c \sin\theta_{ij,k} + b_{ij}^c \cos\theta_{ij,k} + g_{ij}^c \theta_{ij,k} \cos\theta_{ij,k}\right) v_{i,k} v_{j,k} \tag{3l}$$

$$b_{ij}^Q = \left(-g_{ij}^c \alpha_{ij,k}^0 + b_{ij}^c \beta_{ij,k}^0\right) + \left(-g_{ij}^c \alpha_{ij,k}^1 + b_{ij}^c \beta_{ij,k}^1\right)\theta_{ij,k} + b_{ij}^c \beta_{ij,k}^2 \theta_{ij,k}^2 = -g_{ij}^c \sin\theta_{ij,k} + b_{ij}^c \cos\theta_{ij,k} \tag{3m}$$

$$g_{ij}^Q = \left(g_{ij}^c \alpha_{ij,k}^1 - b_{ij}^c \beta_{ij,k}^1\right) v_{i,k} v_{j,k} = \left(g_{ij}^c \cos\theta_{ij,k} + b_{ij}^c \sin\theta_{ij,k} - b_{ij}^c \theta_{ij,k} \cos\theta_{ij,k}\right) v_{i,k} v_{j,k} \tag{3n}$$

$$b_{ij}^{Ploss} = g_{ij}^c \beta_{ij,k}^2 v_{i,k} v_{j,k} = \left(-\frac{1}{2} g_{ij}^c \cos\theta_{ij,k}\right) v_{i,k} v_{j,k}, \ g_{ij}^{Qloss} = -b_{ij}^c \beta_{ij,k}^2 v_{i,k} v_{j,k} = -\left(-\frac{1}{2} b_{ij}^c \cos\theta_{ij,k}\right) v_{i,k} v_{j,k} \tag{3o}$$

By applying a second-order Taylor series expansion to the trigonometric functions in the power flow equations as shown in (2f)–(2i), the bilinear and quadratic terms are retained, yielding (3i) and(3j). In (3k)–(3o), all parameters are constants. Compared with (2f)–(2i), the inherent nonlinearity of the equations is partially mitigated, thereby reducing the computational burden. However, to achieve a fully convex approximation of the power flow equations, further convex relaxation is required.

## 2.4 SOCR for the power flow equations

The voltage bilinear terms and quadratic angle terms can accurately capture the nonlinear characteristics associated with network losses. Therefore, when performing convex relaxation, it is essential to enhance the accuracy of the convex approximation as much as possible. Additionally, the corresponding research [18] indicated that representing the state space of the power flow equations using $(v^2,\theta)$ can effectively capture their nonlinear features without compromising accuracy. In this study, the $(v^2,\theta)$ state space is applied. By introducing relaxation variables $\phi_{ij}^v$ and $\phi_{ij}^\theta$, the voltage bilinear terms and the quadratic angle terms are relaxed into second-order cone constraints, yielding a fully convex approximation of the power flow equations, as shown in below:

$$\phi_{ij}^v = v_i v_j, \quad \phi_{ij}^\theta = \theta_{ij}^2 \tag{4a}$$

The constraints (4a) remain their non-convex feature. Strategies to address this non-convexity mainly include piecewise linearization and convex relaxation [21]. In this study, the latter method is employed and clarified the physical significance of the relaxation. By replacing constraints (4a) with the following inequalities as shown in (4b), the treatment of such SOCR effectively eliminates the non-convexity in the OPF model.

$$\left\| \begin{matrix} 2\phi_{ij}^v \\ v_i^2 - v_j^2 \end{matrix} \right\|_2 \le v_i^2 + v_j^2, \quad \phi_{ij}^v \ge 0, \quad \phi_{ij}^\theta \ge \theta_{ij}^2 \tag{4b}$$

It is known that convex relaxation broadens the feasible region of the original optimization model. The corresponding feasible region transitions from a surface or curve to an entire area enclosed by these surfaces or curves. Consequently, the relaxed power flow equations can be derived as follows:

$$P_{ij,k} = g_{ij}^f v_i^2 - g_{ij,k}^P \phi_{ij}^v - b_{ij,k}^P (\theta_{ij} - \theta_{ij,k}) - b_{ij,k}^{Ploss}(\phi_{ij}^\theta - \theta_{ij,k}^2) \tag{4c}$$

$$Q_{ij,k} = -b_{ij}^f v_i^2 + b_{ij,k}^Q \phi_{ij}^v - g_{ij,k}^Q (\theta_{ij} - \theta_{ij,k}) - g_{ij,k}^{Qloss}(\phi_{ij}^\theta - \theta_{ij,k}^2) \tag{4d}$$

To better illustrate the errors introduced by the SOCR, the corresponding OPF model derived from the second-order Taylor series expansion with $v$ and $\theta$ treated as independent variables, can be expressed as follows:

$$\underset{P_{ig},Q_{ig},v_i^2,\theta_i}{\text{minimize}} \sum_{ig \in \mathcal{N}_{ng}} f_{ig}(P_{ig}) \tag{4e}$$

subject to

$$\underline{P}_{ig} \le P_{ig} \le \overline{P}_{ig} : \tau_{ig}^{Pl}, \tau_{ig}^{Pu}, \quad \underline{Q}_{ig} \le Q_{ig} \le \overline{Q}_{ig} : \tau_{ig}^{Ql}, \tau_{ig}^{Qu}, \quad \forall ig \in \mathcal{N}_{ng} \tag{4f}$$

$$\underline{v}_i \le v_i \le \overline{v}_i : \tau_{ig}^{vl}, \tau_{ig}^{vu}, \quad \underline{\theta}_{ij} \le \theta_i - \theta_j \le \overline{\theta}_{ij} : \tau_{ig}^{\theta l}, \tau_{ig}^{\theta u}, \forall i \in \mathcal{N}, \forall ij \in \mathcal{E} \tag{4g}$$

$$\sum_{ig \in \mathcal{N}_{ng}} P_{ig} - P_{iD} - g_i^{sh} v_i^2 = \sum_{ij \in \mathcal{E}_i} P_{ij,k} + \sum_{ji \in \mathcal{E}_i} P_{ji,k} : \lambda_i^P, \quad \forall i \in \mathcal{N} \tag{4h}$$

$$\sum_{ig \in \mathcal{N}_{ng}} Q_{ig} - Q_{iD} + b_i^{sh} v_i^2 = \sum_{ij \in \mathcal{E}_i} Q_{ij,k} + \sum_{ji \in \mathcal{E}_i} Q_{ji,k} : \lambda_i^Q, \quad \forall i \in \mathcal{N} \tag{4i}$$

$$P_{ij,k} = g_{ij}^f v_i^2 - g_{ij,k}^P \phi_{ij}^v - b_{ij,k}^P (\theta_{ij} - \theta_{ij,k}) - b_{ij,k}^{Ploss}(\phi_{ij}^\theta - \theta_{ij,k}^2) : \lambda_{ij}^{Pij}, \quad \forall i,j \in \mathcal{N}, \forall ij \in \mathcal{E} \tag{4j}$$

$$P_{ji,k} = g_{ij}^t v_j^2 - g_{ji,k}^P \phi_{ij}^v - b_{ji,k}^P (\theta_{ji} - \theta_{ji,k}) - b_{ji,k}^{Ploss}(\phi_{ij}^\theta - \theta_{ji,k}^2) : \lambda_{ji}^{Pji}, \quad \forall i,j \in \mathcal{N}, \forall ji \in \mathcal{E} \tag{4k}$$

$$Q_{ij,k} = -b_{ij}^f v_i^2 + b_{ij,k}^Q \phi_{ij}^v - g_{ij,k}^Q (\theta_{ij} - \theta_{ij,k}) - g_{ij,k}^{Qloss}(\phi_{ij}^\theta - \theta_{ij,k}^2) : \lambda_{ij}^{Qij}, \forall i,j \in \mathcal{N}, \forall ij \in \mathcal{E} \tag{4l}$$

$$Q_{ji,k} = -b_{ij}^t v_j^2 + b_{ji,k}^Q \phi_{ij}^v - g_{ji,k}^Q (\theta_{ji} - \theta_{ji,k}) - g_{ji,k}^{Qloss}(\phi_{ij}^\theta - \theta_{ji,k}^2) : \lambda_{ji}^{Qji}, \forall i,j \in \mathcal{N}, \forall ji \in \mathcal{E} \tag{4m}$$

$$\left\| \begin{matrix} 2\phi_{ij}^v \\ v_i^2 - v_j^2 \end{matrix} \right\|_2 \le v_i^2 + v_j^2 : \varphi_{ij}^v, \quad \phi_{ij}^v \ge 0 : \varphi_{ij}^{vo}, \phi_{ij}^\theta \ge \theta_{ij}^2 : \varphi_{ij}^\theta, \quad \forall i,j \in \mathcal{N}, \forall ij \in \mathcal{E} \tag{4n}$$

$$P_{ij,k}^2 + Q_{ij,k}^2 \le \overline{S}_{ij}^2 : \gamma_{ij}^f, P_{ji,k}^2 + Q_{ji,k}^2 \le \overline{S}_{ij}^2 : \gamma_{ij}^t, \quad \forall ij \in \mathcal{E} \tag{4o}$$

Similar to the original AC OPF model, the objective function (4e) represents the quadratic fossil fuel cost of thermal power unit. Constraints (4f) - (4g) delineate the physical operational constraints of the power grid, while constraints (4h) and (4i) specify the bus power balance constraints. Constraints (4j) - (4n) represent the branch power flow equations after the second-order cone convex approximation, and constraints (4o) describe the branch flow limits in second-order cone form. The variables $\tau_{ig}^{Pl}, \tau_{ig}^{Pu}, \tau_{ig}^{Ql}, \tau_{ig}^{Qu}, \tau_{ig}^{vl}, \tau_{ig}^{vu}, \tau_{ig}^{\theta l}$, and $\tau_{ig}^{\theta u}$ correspond to the dual variables for the physical operational constraints of power grid. The variables $\lambda_i^P, \lambda_i^Q, \lambda_{ij}^{Pij}, \lambda_{ji}^{Pji}, \lambda_{ij}^{Qij}$, and $\lambda_{ji}^{Qji}$ denote the dual variables for the bus power balance and branch flow constraints. Without loss of generality, $\lambda_j^P$ and $\lambda_j^Q$ represent the dual variables for bus $j$. Lastly, $\gamma_{ij}^f$ and $\gamma_{ij}^t$ represent the dual variables for the branch flow limit constraints, while $\varphi_{ij}^v, \varphi_{ij}^{vo}$, and $\varphi_{ij}^\theta$ correspond to the dual variables for the relaxation constraints. To facilitate subsequent problem analysis, the aforementioned model is designated as a SOCR-based Approximate OPF model, hereafter referred to as SOCA-OPF.

**Remark 1:** The Lagrangian function of the SOCA-OPF model can be expressed $\mathcal{L}(P_{ig}, Q_{ig}, v_i^2, \phi_{ij}^v, \theta_i, \varphi_{ij}^\theta)$.

**Remark 2:**
(1) $\lambda_i^P, \lambda_j^P$ and $\lambda_i^Q, \lambda_j^Q$ represent the marginal prices for active and reactive consumption of branch $ij$, respectively.
(2) The summation of $\lambda_i^P$ and $\lambda_j^P$ is usually positive and the value of $|\lambda_i^P + \lambda_j^P|$ is much greater than $|\lambda_i^Q + \lambda_j^Q|$.
(3) In general, for transmission elements, with the exception of special transformers, the conductance $g_{ij}$ is greater than zero, while the susceptance $b_{ij}$ is less than zero.

**Remark 3:** The branch flow limit constraints of most transmission lines are typically non-binding.

**Proposition 1:** In the SOCA-OPF model, the constraints (4j) - (4n) are equivalent to constraints (4a) under the following conditions:

$$-b_{ij,k}^{Ploss}(\lambda_i^P + \lambda_j^P) - g_{ij,k}^{Qloss}(\lambda_i^Q + \lambda_j^Q) + \gamma_{ij}^f(2P_{ij,k}b_{ij,k}^{Ploss} + 2Q_{ij,k}g_{ij,k}^{Qloss}) + \gamma_{ij}^t(2P_{ji,k}b_{ji,k}^{Ploss} + 2Q_{ji,k}g_{ji,k}^{Qloss}) > 0 \quad (5a)$$

**Proof:**

According to the Karush-Kuhn-Tuchker (KKT) conditions, the optimal solution of the SOCA-OPF model satisfies the following:

$$\frac{\partial \mathcal{L}}{\partial \phi_{ij}^\theta} = -\varphi_{ij}^\theta - \left(b_{ij,k}^{Ploss}\lambda_i^P + b_{ji,k}^{Ploss}\lambda_j^P\right) - \left(g_{ij,k}^{Qloss}\lambda_i^Q + g_{ji,k}^{Qloss}\lambda_j^Q\right) \\ -\gamma_{ij}^f\left(2P_{ij,k}\left(-b_{ij,k}^{Ploss}\right) + 2Q_{ij,k}\left(-g_{ij,k}^{Qloss}\right)\right) - \gamma_{ij}^t\left(2P_{ji,k}\left(-b_{ji,k}^{Ploss}\right) + 2Q_{ji,k}\left(-g_{ji,k}^{Qloss}\right)\right) = 0 \quad (5b)$$

Based on (3o), and *Remark* 2(3), it can be concluded that:

$$b_{ij}^{Ploss} = b_{ji}^{Ploss} = \left(-\frac{1}{2}g_{ij}^c \cos\theta_{ij,k}\right)v_{i,k}v_{j,k} < 0, g_{ij}^{Qloss} = g_{ji}^{Qloss} = -\left(-\frac{1}{2}b_{ij}^c \cos\theta_{ij,k}\right)v_{i,k}v_{j,k} < 0 \quad (5c)$$

Therefore, (5b) can be rewritten as:

$$\frac{\partial \mathcal{L}}{\partial \phi_{ij}^\theta} = -\varphi_{ij}^\theta - b_{ij,k}^{Ploss}(\lambda_i^P + \lambda_j^P) - g_{ij,k}^{Qloss}(\lambda_i^Q + \lambda_j^Q) \\ +\gamma_{ij}^f\left(2P_{ij,k}b_{ij,k}^{Ploss} + 2Q_{ij,k}g_{ij,k}^{Qloss}\right) + \gamma_{ij}^t\left(2P_{ji,k}b_{ji,k}^{Ploss} + 2Q_{ji,k}g_{ji,k}^{Qloss}\right) = 0 \quad (5d)$$

Based on *Remark* 2(1)(2) *and Remark* 3, it can be derived that:

$$\varphi_{ij}^\theta = -b_{ij,k}^{Ploss}(\lambda_i^P + \lambda_j^P) - g_{ij,k}^{Qloss}(\lambda_i^Q + \lambda_j^Q) \\ +\gamma_{ij}^f\left(2P_{ij,k}b_{ij,k}^{Ploss} + 2Q_{ij,k}g_{ij,k}^{Qloss}\right) + \gamma_{ij}^t\left(2P_{ji,k}b_{ji,k}^{Ploss} + 2Q_{ji,k}g_{ji,k}^{Qloss}\right) > 0 \quad (5e)$$

According to complementary slackness condition, it holds for optimal solution of SOCA-OPF model:

$$\varphi_{ij}^{\theta}(\phi_{ij}^{\theta} - \theta_{ij}^2) = 0 \tag{5f}$$

Based on (5d) and (5e), in the SOCA-OPF model, the equality $\phi_{ij}^{\theta} = \theta_{ij}^2$ holds true for the majority of branches. Similarly, for $\varphi_{ij}^{v}$, applying the same derivation process yields the following:

$$\phi_{ij}^{v} = v_i v_j \tag{5g}$$

When condition (5a) is satisfied, the constraints (4n) of the SOCA-OPF model are equivalent to the constraints (4a). If the constraints (4a) are not met, fictitious active and reactive losses will be introduced at both ends of branch $ij$. When the marginal price is positive, this typically leads to increased operational costs. Additionally, based on the physical characteristics of the marginal prices in practical power systems, it is generally observed that most systems can satisfy condition (5a). Failure to meet condition (5a) may result in relaxation errors. To address this issue, the next section will propose a rolling cutting plane technique to mitigate such convex errors.

## 2.5 Linear approximation of branch flow limits

In the SOCA-OPF model, the branch flow limit constraints as shown in (4o) is formulated as second-order cone constraints. However, in practical large-scale power grids, the sheer number of branch flow limit constraints can introduce significant numerical challenges when solving large-scale second-order cone problems, compared to linear constraints. To address this issue, this study employs the piecewise linearization method, as proposed in [17], to approximate the constraints (4o).

The feasible region defined by the constraints (4o) is bounded by a black circle, which can be approximated by a set of linear constraints represented by red line segments as shown in Fig. 4. Given the typical dominance of active power flow over reactive power flow in transmission grids, the shaded area as shown in Fig. 4 represents the regions of line flow constraints that are unlikely to occur and are thus classified as infeasible regions.

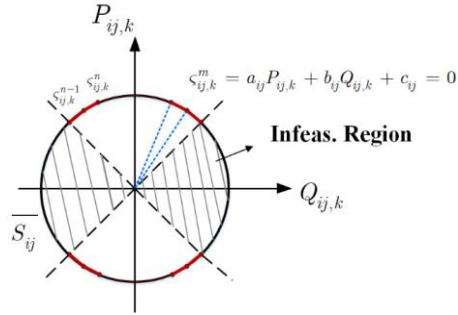

Fig. 4. Linearization of branch flow limit

By focusing on such regions with higher likelihoods of occurrence, the number of linear constraints can be reduced. The more segments used, the smaller the linear error becomes. Consequently, the constraints (4o) can be reformulated as follows:

$$\varsigma_{ij}^{m} \geq 0, m = 1,2,\dots,M \tag{6a}$$

where $\varsigma_{ij}^{m}$ denotes a set of linear constraint combinations that approximately represent the original quadratic power constraints, while $M$ is the total number of segments.

## 2.6 Wind-integrated SOCA-OPF model

Regarding the convexity characteristics of wind generation cost outlined in Section 2.1, a piecewise linearization method is employed through a segmented affine approximation. Additionally, in accordance with the updated grid code provisions [33], wind farms are constrained to operate within a power factor $\cos\phi$ range of 0.95 inductive to 0.95 capacitive. By integrating these operational constraints, the overall wind-integrated SOCA-OPF model is given as follows:

$$\underset{P_{ig},Q_{ig},v_i^2,\theta_i}{\text{minimize}} \sum_{ig\in\mathcal{N}_{ng}} f_{ig}(P_{ig}) + \sum_{iw\in\mathcal{N}_{nw}} \gamma_{iw} \tag{7a}$$

subject to (4h)-(4k),(4n)-(4t),(6a)

$$\gamma_{iw} \geq \eta_l P_{iw,\text{schdule}} + \chi_l, \qquad \forall iw \in \mathcal{N}_{nw}, \forall l \in \mathcal{L} \tag{7b}$$

$$\underline{P}_{\text{wind}} \leq P_{iw,\text{schdule}} \leq \overline{P}_{\text{wind}}, \; Q_{\text{wind}} = \tan\phi \, P_{iw,\text{schdule}}, \quad \forall iw \in \mathcal{N}_{nw} \tag{7c}$$

$$\sum_{ig\in\mathcal{N}_{ng}} P_{ig} + \sum_{iw\in\mathcal{N}_{nw}} P_{\text{schedule}} - P_{iD} - g_i^{\text{sh}} v_i^2 = \sum_{ij\in\mathcal{E}_i} P_{ij,k} + \sum_{ji\in\mathcal{E}_i} P_{ji,k}, \quad \forall i \in \mathcal{N} \tag{7d}$$

$$\sum_{ig\in\mathcal{N}_{ng}} Q_{ig} + \sum_{iw\in\mathcal{N}_{nw}} Q_{\text{wind}} - Q_{iD} + b_i^{\text{sh}} v_i^2 = \sum_{ij\in\mathcal{E}_i} Q_{ij,k} + \sum_{ji\in\mathcal{E}_i} Q_{ji,k}, \quad \forall i \in \mathcal{N} \tag{7e}$$

where $P_{iw,\text{schdule}}$, $\underline{P}_{\text{wind}}$, and $\overline{P}_{\text{wind}}$ denote the scheduled power output of the wind farm and its lower and upper limits, respectively. $Q_{\text{wind}}$ represents the reactive power output of the wind farm. $\gamma_{iw}$, $\eta_l$, and $\chi_l$ are the auxiliary variable, slope, and intercept of the piecewise linear functions.

## 3 Warm-start strategy with rolling cutting plane technique

When constraints (4a) are non-binding, a non-zero relaxation error may occur. To mitigate this convex error, this section presents a rolling cutting plane technique to tighten the convex feasible region. By updating the operational point in each iteration, the algorithm adjusts the convex feasible region to limit the relaxation error, thereby ensuring closer compliance with constraints (4a).

### 3.1 Rolling cutting plane technique for tightening relaxation

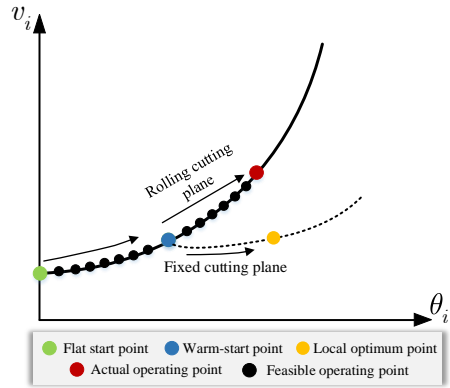

Fig. 5. Schematic diagram for the warm-start strategy with rolling cutting plane technique

To mitigate potential relaxation errors, prior studies have demonstrated significant improvements by adding cutting planes [21]. In this study, a warm-start strategy combined with a proposed rolling cutting plane technique is designed to progressively approach the actual power flow operating point, thereby ensuring the effective reduction of relaxation errors and improving the computational efficiency. In contrast, the conventional fixed cutting plane methodology lacks compatibility with the warm-start mechanism, resulting in deviations in the solution trajectory from the true operational point. Fig. 5. shows the comparative convergence characteristics of various cutting plane strategies. To further limit relaxation errors, the following two rolling cutting planes should be added for the auxiliary variables $\phi_{ij}^{\theta}$ and $\phi_{ij}^{v}$, respectively.

$$\phi_{ij}^{\theta} \leq \left.\frac{\partial \phi_{ij}^{\theta}}{\partial \theta_{ij}}\right|_{\theta_{ij}=\theta_{ij,k}} (\theta_{ij} - \theta_{ij,k}) + \theta_{ij,k}^2 + \Delta_{ij}^{\theta} = 2\theta_{ij,k}\theta_{ij} - \theta_{ij,k}^2 + \Delta_{ij}^{\theta} \tag{8a}$$

$$\phi_{ij}^{v} \geq v_{i,k}v_{j,k} + \left.\frac{\partial \phi_{ij}^{v}}{\partial v_i^2}\right|_{v_i^2=v_{i,k}^2}(v_i^2 - v_{i,k}^2) + \left.\frac{\partial \phi_{ij}^{v}}{\partial v_j^2}\right|_{v_j^2=v_{j,k}^2}(v_j^2 - v_{j,k}^2) - \Delta_{ij}^{v} = \frac{v_{j,k}}{2v_{i,k}}v_i^2 + \frac{v_{i,k}}{2v_{j,k}}v_j^2 - \Delta_{ij}^{v} \tag{8b}$$

where $\Delta_{ij}^{\theta}$ and $\Delta_{ij}^{v}$ represent the upper bounds of the convex relaxation errors for constraints (4a), respectively.

However, as shown in Fig. 5, during the warm-start strategy, the use of fixed cutting plane technique while approaching the true operating point may lead to infeasibility or convergence to a suboptimal solution of inferior quality. In view of this, this study proposes a rolling cutting plane technique to ensure the feasibility of the results.

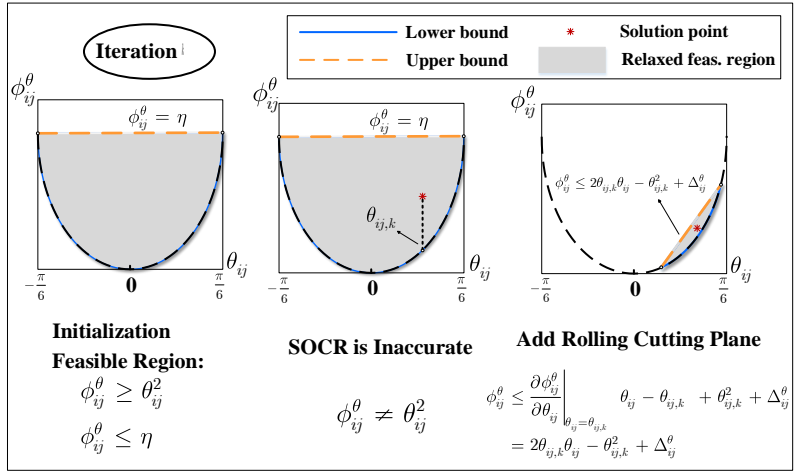

Fig. 6 Schematic diagram of rolling cutting plane technique

In the iterative process, for branches that do not satisfy constraints (4a), the upper bound is estimated using rolling cutting planes formed by the first-order Taylor series expansion. Then, these cutting planes are added to reduce the relaxation errors. As shown in Fig. 6, taking the auxiliary variable $\phi_{ij}^{\theta}$ as an example, the relaxation feasible region is initially defined with a given upper bound. In cases of inaccurate relaxation, the rolling cutting planes technique is leveraged to tighten the upper bound. The corresponding pseudocode for the proposed rolling cutting plane technique is provided in Table I.

Table I Pseudo-code for rolling cutting plane technique

---

**Algorithm 1** Rolling cutting plane technique for wind-integrated SOCA-OPF model

---

**Input:** $\overline{\Delta}_{ij}, \Delta_{ij}^{\theta}, \Delta_{ij}^{v}, \theta_{ij,k}, v_{i,k}$ and $v_{j,k}$
**Initialization:**
1: Feasible Region: $\phi_{ij}^{\theta} \geq \theta_{ij}^{2}, \ \phi_{ij}^{\theta} \leq \eta$.
$$\left\| \begin{matrix} 2\phi_{ij}^{v} \\ v_i^2 - v_j^2 \end{matrix} \right\|_2 \leq v_i^2 + v_j^2, \phi_{ij}^{v} \geq 0$$
2: Solve wind-integrated SOCA-OPF model to obtain $\phi_{ij}^{\theta}, \phi_{ij}^{v}, \theta_{ij}, v_i$, and $v_j$.
3: **if** constraints (3a) and (3b) are satisfied **then**
4:     Return $\theta_{ij,k}, v_{i,k}$, and $v_{j,k}$
5: **else**
6:     **while** $\phi_{ij}^{\theta} + \phi_{ij}^{v} - \theta_{ij,k}^{2} - v_{i,k}v_{j,k} \geq \overline{\Delta}_{ij}$ **do**
7:         Set $\Delta_{ij}^{\theta} = 0.5\Delta_{ij}^{\theta}, \Delta_{ij}^{v} = 0.5\Delta_{ij}^{v}$
8:         Add rolling cutting planes:
$$\phi_{ij}^{\theta} \leq 2\theta_{ij,k}\theta_{ij} - \theta_{ij,k}^{2} + \Delta_{ij}^{\theta}$$
$$\phi_{ij}^{v} \geq \frac{v_{j,k}}{2v_{i,k}}v_i^2 + \frac{v_{i,k}}{2v_{j,k}}v_j^2 - \Delta_{ij}^{v}$$
9:         Solve wind-integrated SOCA-OPF model to obtain $\phi_{ij,k}^{\theta}, \phi_{ij,k}^{v}, \theta_{ij,k}, v_{i,k}$, and $v_{j,k}$
10:    Return $\theta_{ij,k}, v_{i,k}$, and $v_{j,k}$

---

Fig. 7 illustrates the flowchart of the solution process for the wind-integrated SOCA-OPF problem, leveraging enhanced SOCR in conjunction with the proposed rolling cutting plane technique. The initial operating point can be selected as a flat start point $(i.e. v_i = 1, \ \theta_{ij} = 0)$ or a warm-start point obtained from DC OPF solution. In practical applications, the initial point may also be determined based on grid expertise and operational considerations.

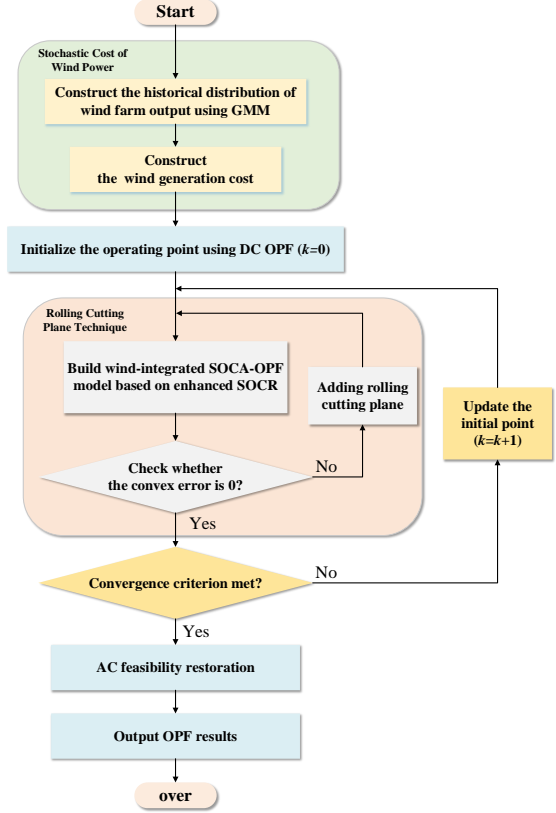

Fig. 7. Flowchart of the solution process for the wind-integrated SOCA-OPF problem

In the warm-start strategy, the convergence criterion is defined based on the errors arising from the approximated power flow equations. Satisfaction of this criterion indicates that the accuracy of the linear approximation is acceptable. In this study, the normalized errors of both branch active and reactive power flows are adopted as the convergence criterion, which is given as follows:

$$\varGamma_{ij} = \frac{\sqrt{P_{ij,k}^2+Q_{ij,k}^2}-\sqrt{P_{ij,e}^2+Q_{ij,e}^2}}{\max\limits_{ij\in\mathcal{E}}\sqrt{P_{ij,e}^2+Q_{ij,e}^2}} \tag{8c}$$

where $P_{ij,e}$ and $Q_{ij,e}$ are accurately calculated from the obtained $v_{i,k}$ and $\theta_{i,k}$. If the condition $\varGamma_{ij} \leq \overline{\varGamma}_{ij}$ holds for all branches, the convergence criterion is satisfied.

## 3.2 AC feasibility restoration

For practical power grids, the obtained solution must ensure AC feasibility. Consequently, it is necessary to perform AC feasibility restoration on the obtained solution. Owing to the convergence criterion given in (8c), the solution exhibits high approximation accuracy. Based on this, the $P_i$ and $Q_i$ of PQ buses, the $P_i$ and $v_i$ of PV buses, and the $v_i$ and $\theta_i$ of Vθ buses are retained. The Newton-Raphson method is then used to perform AC power flow calculation. If the reactive power output of any generator exceeds its limits, the reactive power is fixed, and is the generator is converted to a PQ node for recalculating the AC power flow, ultimately yielding an AC feasible solution.

## 4 Case studies

In this section, the proposed wind-integrated SOCA-OPF model with the rolling cutting plane technique is tested on several IEEE benchmark transmission and distribution networks, with data sourced from MATPOWER. For branches exhibiting zero resistance, a small resistance of $1 \times 10^{-4}$ p.u. is introduced. The value of relaxation error upper bound $\overline{\Delta}_{ij}$ and the convergence criterion limit $\varGamma_{ij}$ are set to $1 \times 10^{-4}$

p.u. and $1 \times 10^{-3}$ p.u., respectively. This study employs wind generation data from the ARPA-E PERFORM [35], with a specific focus on the Black Jack Creek Wind Farm in the United States. The configuration of the selected wind farm comprises 50 wind turbine units, each rated at 4.5 MW, yielding a total installed capacity of 225 MW. The system operates at a power factor of 0.975 ($\cos\phi = 0.975$). For model validation, the complete operational data from the year 2018 is utilized. The GMM is implemented with 12 components ($K = 12$) to accurately represent the characteristics of wind power.

To rigorously validate the robustness of the proposed solution framework, the comprehensive numerical analyses are conducted using two benchmark test systems: the IEEE 118-bus system and the PEGASE 1354-bus system. The methodology is further verified through scalability tests performed on multiple transmission and distribution networks with varying topological complexities. In particular, modified configurations of PEGASE 1354-bus system, incorporating stochastic wind generation, is employed to evaluate the effectiveness of the proposed wind-integrated SOCA-OPF model. The computations are executed on a computer equipped with an AMD Ryzen 5 1600X 6-Core CPU. In this analysis, the following two benchmark methods, denoted as M1 and M2, are compared against the proposed solution framework:

(1) M1 [19]: the linear power flow model with $(v^2, \theta)$ under cold start condition.
(2) M2 [21]: the second-order cone power flow model with $(v^2, \theta)$ under cold start condition.

### 4.1 The IEEE 118-bus system

Simulations are first conducted on the IEEE 118-bus system to evaluate the proposed solution framework. Key performance metrics included the errors in branch active and reactive power losses, errors in branch active and reactive power flows, and errors in voltage magnitude and phase angle. This comparative analysis is essential for assessing the accuracy and reliability of the proposed solution framework in practical applications.

*1) Branch active and reactive power loss error comparison*

The active and reactive power losses in transmission systems exhibit strong nonlinearity. The proposed approximation method demonstrates superior accuracy in estimating these losses, as presented in Fig. 8, which compares the line active and reactive power losses obtained using the proposed method against those from M1 and M2.

From Fig. 8, M1 demonstrates discrepancies due to its neglect of multiple nonlinear terms in the power flow equations. In contrast, M2, which employs SOCR for voltage bilinear terms, improves accuracy, reducing the mean active power loss error to 0.031 p.u. However, notable estimation errors persist in specific line segments, particularly for reactive power. The proposed warm-start strategy combined with rolling cutting plane technique enhances approximation fidelity through successive convex optimization. Numerical tests on the IEEE 118-bus system confirm its effectiveness, maintaining average active power loss errors below $1 \times 10^{-4}$ p.u. and reactive power loss errors under $1 \times 10^{-3}$ p.u. across all operating conditions.

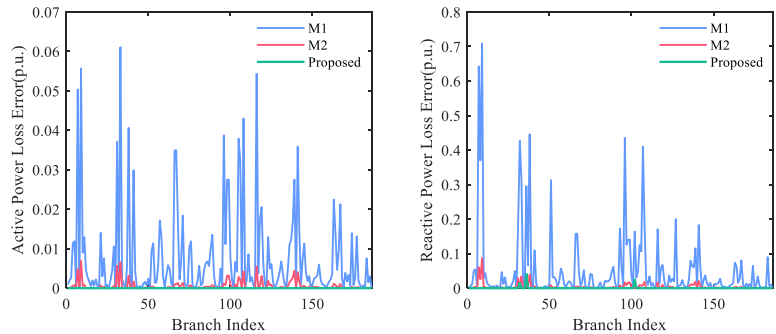

Fig. 8 Comparison of branch active and reactive power loss errors in the IEEE 118-bus system

*2) Branch active and reactive power flow error comparison*

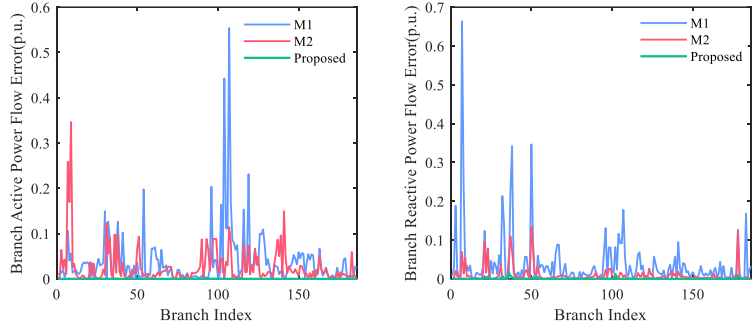

Fig. 9 Comparison of branch active and reactive power flow errors in the IEEE 118-bus system

Fig. 9 demonstrates the branch active and reactive power flow computational errors for the three methods. Consistent with the power loss errors in Fig. 8, the inaccuracies of M1 and M2 propagate to their power flow computation. The proposed warm-start strategy achieves nearly two orders of magnitude improvement in precision. Specifically, M1 yields maximum active and reactive power flow errors of 0.5542 p.u. and 0.6866 p.u., respectively, while M2 reduces these to 0.0322 p.u. (active) and 0.3099 p.u. (reactive). In comparison, the proposed solution framework maintains maximum deviations at merely 0.0012 p.u. (active) and 0.0247 p.u. (reactive).

*3) Bus voltage magnitude and phase angle error comparison*

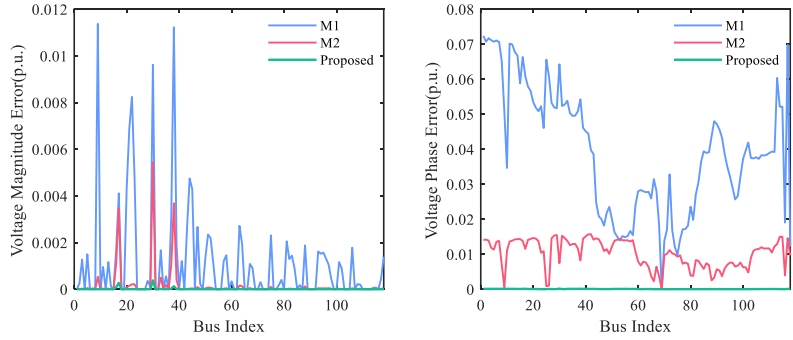

Fig. 10 Comparison of bus voltage magnitude and phase angle errors in the IEEE 118-bus system

To further validate the accuracy, a Newton-Raphson-based AC feasibility restoration is implemented, comparing the three methods against reference power flow solutions. As shown in Fig. 10, M1 exhibits maximum voltage magnitude and phase angle errors of 0.0114 p.u. and 0.072 p.u., respectively, while M2 reduces these to 0.0054 p.u. (voltage magnitude) and 0.0157 p.u. (phase angle). The proposed solution framework achieves AC-feasible solutions with errors reduced by over two orders of magnitude, reaching $3.79 \times 10^{-4}$ p.u. (voltage magnitude) and $7.17 \times 10^{-5}$ p.u. (phase angle).

*4) Performance under varying load conditions*

Comprehensive case studies on the IEEE 118-bus system are conducted under multiple load scaling scenarios, with detailed comparative results documented in Table II.

Table II Accuracy comparison under varying load conditions in the IEEE 118-bus system

| Load Scale | Error in objective function (%) | | | Maximum error in $P_{ij}$(p.u.) | | | Maximum error in $Q_{ij}$(p.u.) | | |
| --- | --- | --- | --- | --- | --- | --- | --- | --- | --- |
| | M1 | M2 | Proposed | M1 | M2 | Proposed | M1 | M2 | Proposed |
| 0.8 | 3.45 | 0.16 | **6.88×10⁻⁴** | 0.34 | 0.03 | **1.05×10⁻³** | 0.44 | 0.33 | **0.03** |
| 0.9 | 3.98 | 0.16 | **7.19×10⁻⁴** | 0.45 | 0.03 | **1.17×10⁻³** | 0.56 | 0.30 | **0.02** |
| 1.0 | 4.48 | 0.14 | **6.43×10⁻⁴** | 0.55 | 0.03 | **1.15×10⁻³** | 0.69 | 0.31 | **0.02** |
| 1.1 | 4.33 | 0.13 | **2.65×10⁻⁴** | 0.60 | 0.03 | **3.02×10⁻³** | 0.74 | 0.31 | **0.07** |
| 1.2 | 3.98 | 0.12 | **2.41×10⁻⁴** | 0.61 | 0.03 | **2.28×10⁻³** | 0.76 | 0.29 | **0.06** |
| 1.3 | 3.78 | 0.11 | **2.31×10⁻⁴** | 0.64 | 0.03 | **7.88×10⁻⁴** | 0.77 | 0.27 | **0.03** |
| 1.4 | 3.66 | 0.11 | **2.16×10⁻⁴** | 0.67 | 0.04 | **9.55×10⁻⁴** | 0.79 | 0.21 | **0.02** |

The proposed solution framework consistently outperforms M1 and M2, maintaining objective function and branch power flow errors 1-2 orders of magnitude lower, as detailed in Table II. This illustrates its robustness in maintaining computational accuracy across diverse operating conditions.

### 4.2 The PEGASE1354-bus system

To validate the extensibility and practical applicability of the proposed solution framework, the comprehensive testing is performed on a large-scale PEGASE 1354-bus power system, with detailed comparative analysis. Owing to space limitations, the simulation comparisons are limited to the bus voltage magnitude and phase angle errors.

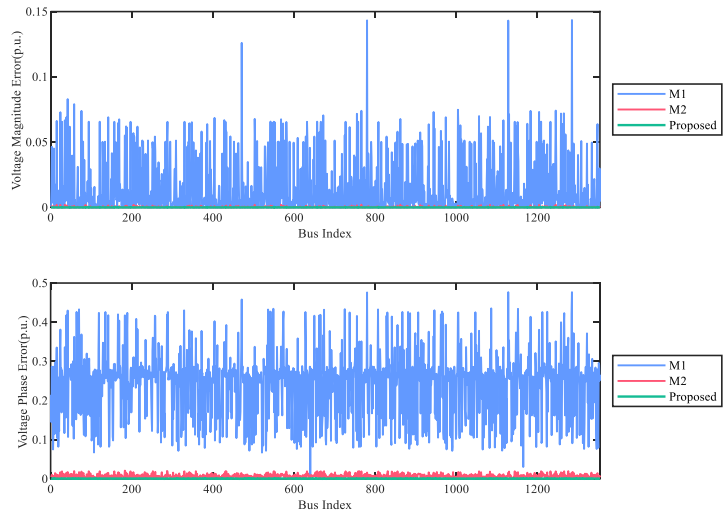

Fig. 11 Comparison of bus voltage magnitude and phase angle errors in the PEGASE 1354-bus system

The proposed solution framework maintains high approximation accuracy in large-scale power systems. As shown in Fig. 11, the post-AC feasibility restoration demonstrates minimal deviations in voltage magnitude and phase angle. For grid-scale systems, the limitations of M1 include the linear power flow formulation's inability to meet accuracy requirements, resulting in substantial errors (>0.05 p.u.) in voltage profiles and angular differences due to inherent nonlinearity approximation deficiencies. Although SOCR in M2 reduce errors by 40-60%, residual discrepancies (>0.02 p.u.) persist due to incomplete capture of operational points, particularly exacerbating reactive power-voltage (QV) issues with over 1000 buses. The proposed solution framework achieves maximum errors of less than 0.001 p.u. on the PEGASE 1354-bus system through warm-start strategy with rolling cutting plane technique, effectively converging to ground-truth power flow solutions within a 0.1% tolerance.

### 4.3 Comparative analysis across other benchmark test systems

To rigorously validate the operational applicability of the proposed solution framework, extensive numerical experiments are conducted on original OPF problems across transmission and distribution (T&D) systems of varying scales, utilizing industry-standard test cases from MATPOWER 7.1. The benchmark suite encompasses 10 representative systems ranging from 14-bus to 2869-bus configurations, systematically evaluating the following metrics: (a) error in objective function, (b) maximum error in voltage magnitude and phase angle, and (c) computational efficiency.

Table III Comparative analysis of accuracy metrics across various transmission systems

| Test case | Error in objective function (%) | | | Maximum error in $v_i$ | | | Maximum error in $\theta_i$ | | | Solution time (Number of iterations) | | |
|---|---|---|---|---|---|---|---|---|---|---|---|---|
| | M1 | M2 | Proposed | M1 | M2 | Proposed | M1 | M2 | Proposed | M1 | M2 | Proposed |
| case 14 | 6.54 | 0.59 | **1.16×10⁻⁴** | 0.007 | 5.4×10⁻⁴ | **2.13×10⁻⁶** | 0.019 | 0.031 | **6.47×10⁻⁶** | 0.06 | 0.16 | 0.36(2) |
| case 30 | 2.15 | 0.48 | **1.09×10⁻⁴** | 0.003 | 0.002 | **1.59×10⁻⁴** | 0.011 | 0.002 | **4.22×10⁻⁵** | 0.06 | 0.25 | 0.58(2) |
| case 118 | 4.48 | 0.14 | **6.43×10⁻⁴** | 0.011 | 0.005 | **3.79×10⁻⁴** | 0.072 | 0.016 | **7.17×10⁻⁵** | 0.08 | 0.33 | 1.17(3) |
| case 300 | 2.77 | 0.06 | **4.92×10⁻⁴** | 0.078 | 0.083 | **0.003** | 0.437 | 0.045 | **0.009** | 0.09 | 0.76 | 1.78(3) |
| case1354pegase | 3.73 | 0.48 | **5.37×10⁻³** | 0.144 | 0.002 | **3.47×10⁻⁴** | 0.477 | 0.021 | **0.0014** | 3.66 | 4.75 | 10.43(3) |
| case2869pegase | 3.59 | 0.27 | **9.93×10⁻³** | 0.087 | 0.015 | **0.0022** | 0.812 | 0.083 | **0.012** | 4.25 | 6.27 | 13.11(2) |

Table III presents a comparative analysis of three methodologies across eight transmission systems of varying scales, focusing on discrepancies in the objective function and errors in voltage magnitude and phase angle relative to AC power flow benchmarks. The results demonstrate that the proposed solution framework achieves orders-of-magnitude improvement in approximation accuracy compared to both M1 and M2. Particularly, in large-scale systems, conventional methods exhibit error magnitudes that could potentially lead to operational misjudgments. In contrast, the proposed solution framework maintains estimation errors within 0.3% across all test cases, satisfying practical engineering accuracy requirements without systematic overestimation or underestimation trends.

Regarding the computational efficiency, while the proposed solution framework exhibits comparable execution times to baseline methods, it preserves linear scalability with system dimensionality. This characteristic ensures that the substantial accuracy improvements are achieved with minimal computational overhead, establishing a favorable accuracy-efficiency trade-off for practical power system optimizations.

Table IV Comparative analysis of performance indicators for distribution systems

| Test case | Maximum error in $v_i$ | | | Maximum error in $\theta_i$ | | | Solution time (Number of iterations) | | |
|---|---|---|---|---|---|---|---|---|---|
| | M1 | M2 | Proposed | M1 | M2 | Proposed | M1 | M2 | Proposed |
| case 33bw | 0.003 | 2.21×10⁻⁵ | **8.05×10⁻¹⁰** | 0.001 | 0.001 | **3.25×10⁻⁷** | 0.01 | 0.01 | 0.17(2) |
| case 33mg | 0.003 | 1.22×10⁻⁵ | **3.53×10⁻⁸** | 0.002 | 8.29×10⁻⁴ | **3.65×10⁻⁷** | 0.03 | 0.15 | 0.23(2) |
| case 69 | 0.004 | 3.07×10⁻⁴ | **2.13×10⁻⁶** | 0.002 | 0.002 | **8.18×10⁻⁷** | 0.04 | 0.22 | 0.39(2) |
| case 141 | 0.002 | 0.152 | **2.12×10⁻⁴** | 3.03×10⁻⁴ | 0.412 | **8.10×10⁻⁶** | 0.09 | 0.41 | 0.52(2) |

Table IV demonstrates the performance of the proposed solution framework in four typical distribution systems from the MATPOWER testbed, where variations in the objective function show limited discrimination potential. It is observed that the proposed solution framework maintains maximum voltage and phase angle deviations below 1×10⁻⁴ p.u., achieving near-exact AC-feasible solutions across all evaluated scenarios. The computational efficiency remains practically viable, with execution times ranging from 0.17 to 0.52 seconds for systems spanning 33-bus to 141-bus configurations.

Notably, the proposed solution framework demonstrates remarkable consistency, preserving solution accuracy irrespective of varying operating conditions or line R/X ratios, and confirms the practical applicability for distribution network optimizations.

### 4.4 Simulations with wind generation cost integrated

In this section, owing to space limitations, the validation of the proposed wind-integrated SOCA-OPF

model is conducted solely through numerical experiments on the PEGASE 1354-bus system. The inherent convex characteristics of wind generation cost modeling facilitate seamless integration with diverse power systems. Therefore, this system verification approach sufficiently demonstrates the applicability and scalability of the proposed solution framework, despite the limited number of test cases.

In the modified PEGASE 1354-bus test system, a wind farm is integrated at Bus 53 with cost coefficients configured as $k_\mathrm{L} = 50, k_\mathrm{H} = 60(\$/MW)$.

Table V Performance evaluation of wind-integrated PEGASE 1354-bus system

| Operating Condition | Maximum Error in $v_i$ | Maximum Error in $\theta_i$ | Solution Time(s) | Total Generation Cost ($) | Fossil Fuel Cost ($) | Wind Generation Cost ($) | Scheduled Wind Farm Power Output (MW) |
|---|---|---|---|---|---|---|---|
| Integrated Bus 53 | 3.11×10⁻⁴ | 1.37×10⁻³ | 10.94 | 78680 | 76355 (97.04%) | 2325.9 (2.96%) | 40.5 |

As shown in Table V, the proposed solution framework maintains high-accuracy solutions in large-scale systems. Additionally, this solution framework extends original OPF formulations by explicitly addressing the characteristics of wind generation cost.

## 5 Conclusion

This study presents a solution framework for wind-integrated SOCA-OPF model, incorporating a warm-start strategy combined with a proposed rolling cutting plane technique to effectively reduce relaxation errors and enhance computational efficiency. Additionally, a data-driven methodology utilizing GMM is employed to statistically characterize historical wind generation data, thereby constructing an analytical expression for the wind generation cost, and introducing into the SOCA-OPF model. Comprehensive numerical simulations conducted on several standard MATPOWER test cases demonstrate the validity and applicability of the proposed solution framework across different system configurations. The experimental results reveal significant improvements in solution accuracy and computational efficiency compared to conventional OPF approaches. Building upon the convexity of the SOCA-OPF model, the proposed solution framework exhibits inherent structural scalability for large-scale power system applications. This fundamental characteristic not only ensures computational tractability but also provides theoretical guarantees for solution feasibility.

## Appendix.

### *GMM fitting for wind power output*

Based on historical operational data from the wind farm, a GMM is employed to fit the PDF of wind power output, with its parameters optimized using the Expectation-Maximization (EM) algorithm. The PDF of the GMM can be expressed as:

$$f(\mathbf{x}) = \sum_{k=1}^{K} \omega_k \mathcal{N}_\mathrm{k}(\mathbf{x}|\boldsymbol{\mu_k}, \boldsymbol{\sigma_k}) \tag{A1}$$

$$\mathcal{N}_\mathrm{k}(\mathbf{x}|\boldsymbol{\mu_k}, \boldsymbol{\sigma_k}) = \frac{\exp\left(-\frac{1}{2}(\mathbf{x}-\boldsymbol{\mu_k})^T(\sigma_k)^{-1}(\mathbf{x}-\boldsymbol{\mu_k})\right)}{\sqrt{(2\pi)^D|\sigma_k^{-1}|}}, \sum_{k=1}^{K} \omega_k = 1, \omega_k > 0 \tag{A2}$$

where $\mathbf{x} = [x_1, x_2, \ldots, x_D]$ denotes a D-dimensional observation vector; $f(\mathbf{x})$ represents the joint PDF; $\omega_k$ is the mixture weight coefficient; $\mathcal{N}_\mathrm{k}(\mathbf{x}|\boldsymbol{\mu_k}, \boldsymbol{\sigma_k})$ denotes the *k*-th multivariate Gaussian component; $\boldsymbol{\mu_k}$ represents the mean vector; $\boldsymbol{\sigma_k}$ is the positive-definite covariance matrix.

Utilizing wind power dataset from the ARPA-E PERFORM program [34], this study selects the Blackjack Creek Wind Farm in the United States as the research case. Under the scheduled active power condition of $P_\mathrm{schedule} = 108\mathrm{MW}$, the estimated PDF of the wind farm power output is fitted, as shown in Fig. 12. The GMM is implemented with 12 components for this analysis.

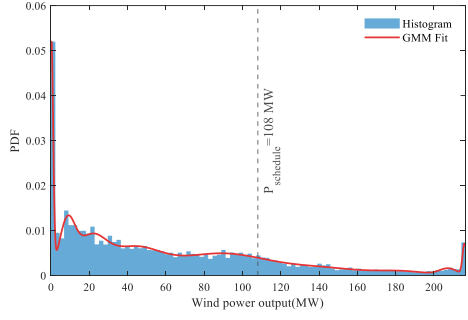

Fig. 12 GMM fitting and historical distribution of wind power output PDF

**Proposition 2:** Let $X$ be a random variable characterized by a GMM, with its PDF denoted by $f_X(x)$. Then, $\int_{\text{dn}}^{\text{up}} v f_X(v) dv = \sum_{k=1}^{K} \omega_k \{\sigma_k^2 [\mathcal{N}_k(\text{dn}) - \mathcal{N}_k(\text{up}) + \mu_k [\Phi_k(\text{up}) - \Phi_k(\text{dn})]]\}$ holds.

**Proof:**

$$\begin{aligned}
\int_{\text{dn}}^{\text{up}} v f_X(v) dv &= \int_{\text{dn}}^{\text{up}} v [\sum_{k=1}^{K} \omega_k \mathcal{N}_k(x|\mu_k, \sigma_k)] dv \\
&= \sum_{k=1}^{K} \omega_k \frac{1}{\sqrt{\pi}} \int_{\text{dn}}^{\text{up}} v e^{-\left(\frac{v-\mu_k}{\sqrt{2}\sigma_k}\right)^2} d\left(\frac{v-\mu_k}{\sqrt{2}\sigma_k}\right) \\
&= \sum_{k=1}^{K} \omega_k \left[\frac{\sqrt{2}\sigma_k}{2\sqrt{\pi}} \int_{\text{dn}}^{\text{up}} e^{-\left(\frac{v-\mu_k}{\sqrt{2}\sigma_k}\right)^2} d\left(\frac{v-\mu_k}{\sqrt{2}\sigma_k}\right)^2 + \frac{\mu_k}{\sqrt{\pi}} \int_{\text{dn}}^{\text{up}} e^{-\left(\frac{v-\mu_k}{\sqrt{2}\sigma_k}\right)^2} d\left(\frac{v-\mu_k}{\sqrt{2}\sigma_k}\right)\right] \\
&= \sum_{k=1}^{K} \omega_k \{\sigma_k^2 [\mathcal{N}_k(\text{dn}) - \mathcal{N}_k(\text{up}) + \mu_k [\Phi_k(\text{up}) - \Phi_k(\text{dn})]]\}
\end{aligned} \quad (A3)$$

# References

[1] J. Carpentier, "Contribution to the Economic Dispatch Problem," Bull. Soc. Franc. Elect., vol. 8, no. 3, pp. 431–447, 1962.

[2] Molzahn D K, Hiskens I A. A survey of relaxations and approximations of the power flow equations[J]. Foundations and Trends® in Electric Energy Systems, 2019, 4(1-2): 1-221.

[3] Torres G L, Quintana V H. An interior-point method for nonlinear optimal power flow using voltage rectangular coordinates[J]. IEEE transactions on Power Systems, 1998, 13(4): 1211-1218.

[4] Lavaei J, Low S H. Zero duality gap in optimal power flow problem[J]. IEEE Transactions on Power systems, 2011, 27(1): 92-107.

[5] Jabr R A, Coonick A H, Cory B J. A primal-dual interior point method for optimal power flow dispatching[J]. IEEE Transactions on Power Systems, 2002, 17(3): 654-662.

[6] Hasan M S, Chowdhury M M U T, Kamalasadan S. Sequential quadratic programming (SQP) based optimal power flow methodologies for electric distribution system with high penetration of DERs[J]. IEEE Transactions on Industry Applications, 2024.

[7] Mhanna S, Mancarella P. An exact sequential linear programming algorithm for the optimal power flow problem[J]. IEEE Transactions on Power Systems, 2021, 37(1): 666-679.

[8] Jabr R A. Radial distribution load flow using conic programming[J]. IEEE transactions on power systems, 2006, 21(3): 1458-1459.

[9] Farivar M, Low S H. Branch flow model: Relaxations and convexification—Part I[J]. IEEE Transactions on Power Systems, 2013, 28(3): 2554-2564.

[10] Coffrin C, Hijazi H L, Van Hentenryck P. The QC relaxation: A theoretical and computational study on optimal power flow[J]. IEEE Transactions on Power Systems, 2015, 31(4): 3008-3018.

[11] Bai X, Wei H, Fujisawa K, et al. Semidefinite programming for optimal power flow problems[J]. International Journal of Electrical Power & Energy Systems, 2008, 30(6-7): 383-392.

[12] Farivar M, Low S H. Branch flow model: Relaxations and convexification—Part I[J]. IEEE Transactions on Power Systems, 2013, 28(3): 2554-2564.

[13] Conejo A J, Aguado J A. Multi-area coordinated decentralized DC optimal power flow[J]. IEEE transactions on power systems, 1998, 13(4): 1272-1278.

[14] Stott B, Alsaç O. Optimal power flow: Basic requirements for real-life problems and their solutions[C]//SEPOPE XII Symposium, Rio de Janeiro, Brazil. 2012, 11: 1-10.

[15] Baker K. Solutions of DC OPF are never AC feasible[C]//Proceedings of the Twelfth ACM International Conference on Future Energy Systems. 2021: 264-268.

[16] Yang Z, Zhong H, Bose A, et al. A linearized OPF model with reactive power and voltage magnitude: A pathway to improve the MW-only DC OPF[J]. IEEE Transactions on Power Systems, 2017, 33(2): 1734-1745.

[17] Yang Z, Zhong H, Xia Q, et al. A novel network model for optimal power flow with reactive power and network losses[J]. Electric Power Systems Research, 2017, 144: 63-71.

[18] Yang Z, Xie K, Yu J, et al. A general formulation of linear power flow models: Basic theory and error analysis[J]. IEEE Transactions on Power Systems, 2018, 34(2): 1315-1324.

[19] Zhang H, Heydt G T, Vittal V, et al. An improved network model for transmission expansion planning considering reactive power and network losses[J]. IEEE Transactions on Power Systems, 2013, 28(3): 3471-3479.

[20] Fan Z, Yang Z, Yu J. Error bound restriction of linear power flow model[J]. IEEE Transactions on Power Systems, 2021, 37(1): 808-811.

[21] Xiao Y, Bie Z, Huang G, et al. Chance-constrained distributional robust optimization based on second-order cone optimal power flow[J]. Power System Technology, 2021, 45(4): 1505-1517.

[22] Li Z, Yu J, Wu Q H. Approximate linear power flow using logarithmic transform of voltage magnitudes with reactive power and transmission loss consideration[J]. IEEE Transactions on Power Systems, 2017, 33(4): 4593-4603.

[23] Fan Z, Yang Z, Yu J, et al. Component Analysis and Accuracy Improvement of Linear Power Flow Equation Based on Legendre Polynomial Expansion[J]. IEEE Transactions on Power Systems, 2022, 38(5): 4617-4627.

[24] Long J, Yang Z, Zhao J, et al. Modular linear power flow model against large fluctuations[J]. IEEE Transactions on Power Systems, 2023, 39(1): 402-415.

[25] Louca R, Bitar E. Robust AC optimal power flow[J]. IEEE Transactions on Power Systems, 2018, 34(3): 1669-1681.

[26] Bai X, Qu L, Qiao W. Robust AC optimal power flow for power networks with wind power generation[J]. IEEE Transactions on Power Systems, 2015, 31(5): 4163-4164.

[27] Yong T, Lasseter R H. Stochastic optimal power flow: formulation and solution[C]//2000 Power Engineering Society Summer Meeting (Cat. No. 00CH37134). IEEE, 2000, 1: 237-242.

[28] Bazrafshan M, Gatsis N. Decentralized stochastic optimal power flow in radial networks with distributed generation[J]. IEEE Transactions on Smart Grid, 2016, 8(2): 787-801.

[29] Miranda V, Hang P S. Economic dispatch model with fuzzy wind constraints and attitudes of dispatchers[J]. IEEE Transactions on power systems, 2005, 20(4): 2143-2145.

[30] Hetzer J, David C Y, Bhattarai K. An economic dispatch model incorporating wind power[J]. IEEE Transactions on energy conversion, 2008, 23(2): 603-611.

[31] Shi L, Wang C, Yao L, et al. Optimal power flow solution incorporating wind power[J]. IEEE Systems Journal, 2011, 6(2): 233-241.

[32] National Development and Reform Commission. (2019, May 25). Notice on improving the on-grid electricity price policy for wind power. NDRC Website. https://www.gov.cn/xinwen/2019-05/25/content_5394615.htm.

[33] Tsili M, Papathanassiou S. A review of grid code technical requirements for wind farms[J]. IET Renewable power generation, 2009, 3(3): 308-332.

[34] Sergi, Brian, Feng, Cong, Zhang, Flora, Hodge, Bri-Mathias, Ring-Jarvi, Ross, Bryce, Richard, Doubleday, Kate, Rose, Megan, Buster, Grant, & Rossol, Michael. ARPA-E PERFORM datasets. United States. https://dx.doi.org/10.25984/1891136.